\documentclass[12pt]{article}
\begin{document}
\title{GRAVITATION AS A ``WEAK ELECTROMAGNETISM'' AND OTHER ISSUES}
\author{B.G. Sidharth\\
International Institute for Applicable Mathematics \& Information Sciences\\
Hyderabad (India) \& Udine (Italy)\\
B.M. Birla Science Centre, Adarsh Nagar, Hyderabad - 500 063 (India)}
\date{}
\maketitle
\begin{abstract}
In the context of a Planck scale underpinning for the universe, we argue that both gravitation and electromagnetism can be characterized in a unified way, in a Sakharov like description. We also consider the issue of observed gamma radiation, which recent observations seem to indicate are caused by the annihilation of exotic $Mev$ particles, as indeed was theoretically predicted earlier. Some other problems are also considered.
\end{abstract} 
\section{Gravitation and Electromagnetism}
It was shown that a pion could be considered to be the lowest energy state of $n \sim 10^{40}$ Planck oscillators, there being $N \sim 10^{80}$ pions in the universe, while the universe itself could be considered to have an underpinning of $nN = \bar {N} \sim 10^{120}$ of these Planck oscillators \cite{fpl,fpl2}. The number $N$ is well known from the time of Eddington, while $\bar N$ features in more recent studies (Cf. also \cite{nottale}). A consequence of this is the set of relations quoted below
\begin{equation}
l = \sqrt{n} l_P, \quad t = \sqrt{n} t_P\label{e1}
\end{equation}
\begin{equation}
R = \sqrt{N} l, \, T = \sqrt{N} t \, R = \bar N^{\frac{1}{2}} l_P, \, T = \bar N^{\frac{1}{2}} t_P\label{e2}
\end{equation}
\begin{equation}
M = Nm = \bar N^{\frac{1}{2}} m_P\label{e3}
\end{equation}
where $R \sim 10^{28}cms, \, T \sim 10^{17}secs,  \, M \sim 10^{55}gms$ are the radius, age and mass of the universe, $l, t\, \mbox{and}\, m$ being the Compton length, Compton time and mass of a typical elementary particle like the pion and finally $l_P , t_P, m_P$ are the Planck length, time and mass. All these relations can be easily verified to be correct.\\
It may be mentioned that the first of equations (\ref{e2}) is the well known Eddington formula. This and similar equations involving $N$ were the basis for the Dirac Cosmology \cite{narlikar} and should be considered in an order of magnitude sense, or as Wheeler puts it \cite{mwt}, the distinction between the masses of the electron and the pion, for example being ignored.\\
Let us use the well known fact \cite{nottale2} that the gravitational energy of an isotropic sphere of $N$ elementary particles of radius $R$ can be obtained from the equation
\begin{equation}
R = \frac{GM}{c^2} = \frac{GNm}{c^2}\label{e4}
\end{equation}
We can see that the mass $M$ or energy of the universe obtained from (\ref{e4}) is in terms of gravitation, while the same mass $M$ (or energy) obtained from (\ref{e3}) and (\ref{e2}) is in terms of the underpinning of Planck oscillators without any reference to gravitation. Infact equating the two we get \cite{bgspreprint}
\begin{equation}
G = \frac{lc^2}{m\sqrt{N}}\label{e5}
\end{equation}
We can easily verify that (\ref{e5}) is indeed correct. Infact (\ref{e5}) was alternatively deduced in a model that correctly predicted a dark energy driven accelerating universe with a small cosmological constant as also deduced all the large number coincidental relations referred to along with the mysterious so called Weinberg formula
\cite{mg8,ijmpa,ijtp}
$$m = \left(\frac{H\hbar^2}{Gc}\right)^{\frac{1}{3}}$$
which was otherwise considered to be an inexplicable accident.\\
We note that (\ref{e5}) can be rewritten as the well known electromagnetism-gravitation ratio
\begin{equation}
\frac{e^2}{Gm^2} \sim \sqrt{N}\label{e6}
\end{equation}
Equation (\ref{e6}) has also been considered to be one of the coincidental large number relations. We can appreciate the significance of (\ref{e5}) or (\ref{e6}) if, following Hayakawa \cite{hayakawa} we equate the excess of electrical potential energy of the electrons in the universe due to the fluctuation in the particle number $\sim \sqrt{N},$ with the inertial energy of the elementary particle, viz.,
$$\frac{\sqrt{N} e^2}{R} \approx mc^2$$
This leads to (\ref{e6}) or (\ref{e5}). Infact it was pointed out \cite{bgsx} that once (\ref{e6}) was deduced rather than taken as an adhoc coincidence, this meant that already it pointed towards an unification of electromagnetism and gravitation. Equations (\ref{e5}) or (\ref{e6}) show that gravitation appears as the excess or residual energy over $N$ particles rather on the lines of Sakharov's analysis using the Quantum vaccuum and the Planck scale, as indeed has been the rationale above \cite{mwt,sakharov}.\\
The question that arises is, can we similarly consider the electromagnetic interaction between elementary particles to be the residual energy of the underpinning Planck oscillators, between elementary particles. In other words in (\ref{e5}) if we replace $N$ by a number $P$ which is $\sim 0(1)$ then we should get instead of $Gm^2$ the gravitational coupling $e^2$ the electromagnetic coupling. Infact we get
\begin{equation}
e^2 \approx lmc^2\label{e7}
\end{equation}
which is indeed true.\\
We can further support the above characterization in equations (\ref{e5}) and (\ref{e6}) of gravitation as a form of ``weak electromagnetism'' (or ``weak electric force'') or electromagnetism as a form of strong gravitation as follows: (It must be borne in mind that the terms weak electromagnetism and strong gravitation were used several years ago in different contexts). Firstly we observe that an equation like (\ref{e4}) with a numerical factor 2 on the right side (which in the large number context is not important) gives the Schwarschild radius of a black hole of mass $M$. If $Gm^2, M$ for the moment being replaced by $m$, is substituted by $e^2$ in (\ref{e4}), then we should get the corresponding ``Schwarschild radius'' for electromagnetism treated as strong gravitation. Indeed we then get (\ref{e7}) giving the Compton wavelength for the mass $m$. In other words the Compton wavelength shows up as a non gravitational but rather electromagnetic Schwarschild radius on the scale of elementary particles.\\
Let us now consider the temperature and life time of a black hole in the context of the Hawking-Beckenstein radiation. In the usual theory we have \cite{ruffini} in standard notation
\begin{equation}
T = \frac{\hbar c^3}{8\pi Gkm}\label{e8}
\end{equation}
\begin{equation}
\frac{dm}{dt} = -\frac{\beta}{m^2},\label{e9}
\end{equation}
where $\beta$ is given by
$$\beta = \frac{\hbar c^4}{(30.8)^3 \pi G^2}$$
This leads to the usual black hole life time given by
\begin{equation}
t = \frac{1}{3\beta} m^3 = 8.4 \times 10^{-24} m^3 secs,\label{e10}
\end{equation}
If now we carry out the substitution $Gm^2 \to e^2$ in the above we have instead of (\ref{e8}), the relation
\begin{equation}
kT \sim mc^2\label{e11}
\end{equation}
Equation (\ref{e11}) is the well known relation expressing the Hagedorn temperature of elementary particles \cite{sivaram}. Similarly instead of (\ref{e9}) we will get
$$\frac{dm}{dt} = -\frac{\hbar c^4}{\Theta^3 e^4} m^2, \, \Theta^3 = (30.8)^3 \pi$$
Whence we get for the life time 
\begin{equation}
\frac{\hbar c^4}{\Theta^3 e^4} t = \frac{1}{m}\label{e12}
\end{equation}
From (\ref{e12}) we get, for the pion, a life time
$$t \sim 10^{-23}secs,$$
which is the pion Compton time.\\
Thus for elementary particles, working within the context of gravitational theory, but with a scaled up coupling constant, we get the meaningful relations (\ref{e7}), (\ref{e12}) and (\ref{e11}) giving the Compton length and Compton time as also the Hagedorn temperature as the analogues of the Schwarzschild radius, radiation life time and black hole temperature obtained with the usual Gravitational coupling constant.
\section{Discussion}
1. The role of the Planck scale in Quantum Gravity considerations is well known. We reiterate that what has been done is that the same reasoning used in the theory of black holes within a purely gravitational framework can be extended to electromagnetic considerations, and then this leads to the Compton scale of elementary particles. In this sense, there is just a rescaling.\\
2. The Planck scale considerations, as is well known lead to a modification of the Uncertainty Principle (Cf. \cite{garay,mup} and several references therein). There is now, in addition to the usual Heisenberg Uncertainty term, an additional term given by
\begin{equation}
\Delta x = l^2_P \frac{\Delta p}{\hbar}\label{e13}
\end{equation}
As the Uncertainty in the momentum $\Delta p \sim \sqrt{n} m_P$, given the fact that as pointed out in the beginning there are $n$ Planck oscillators defining a typical elementary particle, we have from (\ref{e13})
$$l = l^2_P \sqrt{n} \frac{m_P c}{\hbar} = \sqrt{n} l_P$$
which is just (\ref{e1}). So the modification of the Uncertainty relation due to Planck scale considerations lead to the Compton scale.\\
3. Already we have referred to Sakharov's formulation of gravitation in terms of the background Zero Point Field (or Quantum vaccuum) (Cf.ref.\cite{mwt,sakharov}). In this context let us recapitulate the following well known fact \cite{mwt,bgsx}: Due to the Zero Point oscillators, there is an electromagnetic field density $\Delta B$ over an interval $L$ given by
\begin{equation}
\left(\Delta B\right)^2 \sim \frac{e^2}{L^4}\label{e14}
\end{equation}
So the energy over an extension $L = l$ is given from (\ref{e14}) by $\frac{e^2}{l}$ which is the energy $mc^2$ of the elementary particle itself,
\begin{equation}
\frac{e^2}{l} = mc^2\label{e15}
\end{equation}
If on the other hand we replace in (\ref{e15}) $e^2$ by $Gm^2$, we get, reverting to the length $L$
$$\frac{Gm^2}{L} \approx mc^2$$
whence
\begin{equation}
L \approx \frac{Gm}{c^2}\label{e16}
\end{equation}
(\ref{e16}) shows that we can similarly obtain from the fluctuating background Zero Point Field a black hole, infact a Planck scale black hole, it being well known that a Planck mass is a Schwarzschild black hole at the Planck scale (Cf. also ref.\cite{rosen}). From this point of view, Planck mass particles are created from the fluctuation of the zero Point Field and then lead up to elementary particles as above. In any case, this again brings out the interchangability, $e^2 \to Gm^2$.\\
4. We have seen above how from the background Zero Point Field Planck scale particles can ``condense''. Let us suppose that $n$ such particles are formed. We can then use the well known fact that \cite{moller} for a collection of ultra relativistic particles, in this case the Planck oscillators, the various centres of mass form a two dimensional disk of radius $l$ given by
\begin{equation}
l \approx \frac{\beta}{m_e c}\label{e17}
\end{equation}
where in (\ref{e17}) $m_e (\approx m$ in the large number sense) is the electron mass and  $\beta$ is the angular momentum of the system. Further $l$ is such that for distances $r < l$, we encounter negative energies. It will at once be apparent that for an electron, for which $\beta = \frac{\hbar}{2}$, (\ref{e17}) gives the Compton wavelength. We can further characterize (\ref{e17}) as follows: By the definition of the angular momentum of the system of Planck particles moving with relativistic speeds, we have
\begin{equation}
\frac{\hbar}{2} = m_P c \int^l_0 r^2 drd \Theta \, \sim m_P c \sigma l^3 = m_e cl\label{e18}
\end{equation}
In (\ref{e18}) we have used the fact that the disk of mass centres is two dimensional, and $\sigma$ has been inserted to stress the fact that we are dealing with a two dimensional denslity, so that $\sigma$ while being unity has the dimension
$$\left[\frac{1}{L^2}\right]$$
The right side of (\ref{e18}) gives the angular momentum for the electron.
From (\ref{e18}) we get
\begin{equation}
\sigma l^2 m_P = m_e\label{e19}
\end{equation}
which ofcourse is correct.\\
Alternatively from (\ref{e19}) we can recover $n \sim 10^{20}$, in the large number sense, a fact which we encountered earlier. (\ref{e19}) also gives another puzzling coincidence namely
$$l_P = \sigma l^3$$
This is the explanation for the fact that the cube of the electron Compton wavelength numerically equals the Planck length.\\
5. Dirac, and later Wheeler \cite{su3} had considered the possibility that the ``rate of time'' may be different for different interactions. Let us now touch upon this aspect, in the light of the above considerations. Firstly, we observe that there are two basic scales in the universe. The electroweak-strong microscopic scale and the gravitational macroscopic or cosmic scale. The time and length scales for these two broad regions are given by the well known relations
\begin{equation}
R = \sqrt{N} l, \, T = \sqrt{N} \tau\label{e20}
\end{equation}
(Cf. relations (\ref{e2})). However, $c$ is same for both scales. Now if we consider the representation of the Hamiltonian as the time derivative operator we will get
\begin{equation}
H(T) = \frac{d}{dT} = \frac{d}{\sqrt{N} d\tau} = \frac{H(\tau)}{\sqrt{N}}\label{e21}
\end{equation}
wherein we have used (\ref{e20}). $H(T)$ in (\ref{e21}) denotes gravitation represented by the coupling constant $Gm^2$ and $H(\tau )$ in (\ref{e21}) denotes electromagnetism represented by the coupling constant $e^2 e$ and $m$ referring to the same elementary particle. Whence (\ref{e21}) gives
$$\frac{e^2}{Gm^2} \sim \sqrt{N}$$
which infact is the well known supposedly accidental ratio of the coupling constants encountered earlier.\\
Let us now consider the analogue of the microscopic relation,
$$m \frac{l^2}{\tau} = h$$
for the macro or cosmic scale. We then get
\begin{equation}
h gcm^2/sec \to ML^2/T = H \sim 10^{93}\label{e22}
\end{equation}
This equation, is infact perfectly meanlingful because $H$ in (\ref{e22}) is the Godel spin of the universe \cite{su,carn}. Infact (\ref{e22}) leads to
\begin{equation}
R = \frac{H}{Mc}\label{e23}
\end{equation}
(\ref{e22}) and (\ref{e23}) show that the universe itself seems to follow a Quantum Mechanical behaviour with a scaled up Planck constant $H$ as argued previously \cite{bgssu,sqe}.\\
The above considerations can be modelled by the universality and scaling effects of Critical Point Phenomena and the Renormalization Group \cite{davis}: The universe is a coarse grained scaled up version of the micro world, gravitation being the counterpart of electromagnetism given by their mutual scaled ratio.\\
Interestingly if the above considerations are carried over to the Planck scale versus the Compton scale, we can easily verify that there is no new scaled down Planck constant, as for example in (\ref{e22})-- that is the Planck scale considerations remain the same as the Compton scale.\\
6. Considering extremal black holes, it was pointed out by the author \cite{ffp4} that the observed gamma rays could be attributed to particles with mass $\sim 1 Mev$, and with a charge of the order of that of a monopole viz., $1000 e$. It is interesting that recent analysis of observed data by Boehm and co-workers confirms these conclusions \cite{boehm}.\\
Infact there is a spectrum of possibilities as seen below. Let us start with the radius of the Kerr-Newman metric
$$r_+ = \frac{GM}{c^2} + \imath b, b \equiv \left(\frac{G^2Q^2}{c^8} + a^2 - \frac{G^2M^2}{c^4}\right)^{1/2}$$
It can be seen that there is a naked singularity, which however disappears if there is no spin, and moreover 
\begin{equation}
Q \sim Mc^2,\label{e24}
\end{equation}
From (\ref{e24}) it can be seen that apart from the possibility mentioned above, we could have particles with the charge $e$ but mass a thousandth that of an electron or we could have particles with masses that of the neutrino, that is a mass one hundred millionth that of the electron or less. These particles would have a charge one millionth or less the electron charge \cite{bgsffp2} and so on.\\
All these physical microscopic black holes could in principle be produced from the background ZPF, but they would be almost instantaneous particles wilth a negligible Beckenstein Radiation Life Time. However they would annihilate themselves in contact with their anti particles and produce gamma radiation. (Infact it was arguyed that a photon could be considered to be a neutrino anti-neutrino bound state \cite{bgsiced}). This indeed is a conclusion drawn by Boehm and co-workers.\\
7. It was earlier argued \cite{bgsx} that a state can be written as 
\begin{equation}
\psi = \sum_{n} c_n \phi_n,\label{e25}
\end{equation}
in terms of basic states $\phi_n$ which could be eigen states of energy for example, with eigen values $E_n$. It is known that (\ref{e25}) can be written as
\begin{equation}
\psi = \sum_{n} b_n \phi_n\label{e26}
\end{equation}
where $|b_n|^2 = 1 \, \mbox{if} \,E < E_n < E + \Delta, \, \mbox{and} \, = 0$ otherwise under the assumption
\begin{equation}
\overline{(c_n,c_m)} = 0, n \ne m\label{e27}
\end{equation}
(Infact $n$ could stand for not a single state but for a set of states $n_\imath$ and so also $m$). Here the bar denotes a time average over a suitable interval. This is the well known Random Phase Axiom and arises due to the total randomness amongst the phases $c_n$. Also the expectation value of any operator $0$ is given by
\begin{equation}
< 0 > = \sum_{n} |b_n|^2 (\phi_n, 0 \phi_n)/\sum_{n}|b_n|^2\label{e28}
\end{equation}
(\ref{e26}) and (\ref{e28}) show that effectively we have incoherent states $\phi_1,\phi_2,\cdots$ once averages over time intervals for the phases $c_n$ in (\ref{e27}) vanish owing to their relative randomness. In the light of the preceding discussion of random fluctuations, we can interpret all this meaningfully: We can identify $\phi_n$ with the ZPF. The time averages are the zitterbewegung averages over intervals $\sim \frac{\hbar}{mc^2}$. We then get disconnected or incoherent particles from a single background of vaccuum fluctuations exactly as before. The incoherence arises because of the well known random phase relation (\ref{e27}) that is after averaging over the suitable interval.\\
Infact this can be seen more explicitly as follo9ws. We consider now for simplicity the free particle Dirac equation. The solutions are of the type,
\begin{equation}
\psi = \psi_A + \psi_S\label{e29}
\end{equation}
where
$$\psi_A = e^{\frac{1}{\hbar}Et} \left(\begin{array}{ll}
0\\ 0\\ 1\\ 0\end{array} \right) \, \mbox{or} \, e^{\frac{1}{\hbar}Et} \left(\begin{array}{ll} 0\\0\\0\\1\end{array}\right) \, \mbox{and}$$
\begin{equation}
\psi_S = e^{-\frac{1}{\hbar}Et} \left(\begin{array}{ll}1\\0\\0\\0\end{array}\right) \, \mbox{or} \, e^{-\frac{1}{\hbar}Et} \left(\begin{array}{ll} 0\\1\\0\\0\end{array}\right)\label{e30}
\end{equation}
A typical solution would be a super position of all such solutions. From (\ref{e29}) and (\ref{e30}) the probability of finding the particle in a small volume is given by terms like 
\begin{equation}
|\psi_A + \psi_S|^2 = |\psi_A|^2 + |\psi_S|^2 + (\psi_A \psi^*_S + \psi_S \psi^*_A) \cdots \label{e31}
\end{equation}
The point is that in (\ref{e31}) there are interference terms-- terms corresponding to zitterbewegung in the Dirac electron theory. These interference terms vanish, as in (\ref{e27}), when suitable averages are taken, over the Compton scale, and then we are left with terms that contain no interference, like the first two terms on the right side of equation (\ref{e27}). This corresponds to the equation (\ref{e26}).\\
8. We may finally observe that at the Compton scale we have the equation 
\begin{equation}
Gm^2_P \sim e^2\label{e32}
\end{equation}
This relation can in the light of earlier remarks be construed to mean that at the Planck scale, there is only one interaction or energy, which could be taken to be electromagnetic.

\end{document}